# Electronic Structure of Novel Cation-Radical Salts in High Magnetic Fields

J. S. BROOKS,[a*] L. BALICAS,[a] K. STORR,[a] B.H. WARD,[a+]
S. UJI,[b] T. TERASHIMA,[b] C. TERAKURA,[b] J.A. SCHLUETER,[c]
R.W. WINTER,[d] J. MOHTASHAM,[d] G.L. GARD,[d]
G.C. PAPAVASSILIOU,[e] M. TOKUMOTO[f]

[a]NHMFL-Physics, Florida State Univ., Tallahassee FL USA; [b]Tsukuba Magnet Lab-NRIM, Tsukuba, Japan; [c]Chem. and Mat. Sci. Div., Argonne Nat. Lab., Argonne IL USA.; [d]Dept. of Chem., Portland State Univ., Portland OR USA; [e]Theoretical and Physical Chemistry Inst., NHRF., Athens, Greece; [f]Electrotechnical Lab., Tsukuba, Japan.

Abstract  Two organic conducting materials, where unusual aspects of their composition play important roles, are explored: $\beta''$-(BEDT-TTF)$_2$SF$_5$XSO$_3$ which exhibits superconductivity, or a metal-insulator transition (for $X$=CH$_2$CF$_2$ or CHF respectively), and $\tau$-(P-$S, S$-DMEDT-TTF)$_2$ (AuBr$_2$) (AuBr$_2$)$_y$ which exhibits a large, hysteretic, negative magnetoresistance. Detailed angular dependent magnetoresistance studies that allow a tomographic view of the electronic structure of these materials with increasing magnetic fields are presented.

Keywords   organic conductors, high magnetic fields, electronic structure tomography

INTRODUCTION

We consider the electrical transport properties of two very unique organic conductor structures at low temperatures, in the presence of tilted magnetic fields. The first is $\beta''$(also $\beta'$) -(BEDT-TTF)$_2$SF$_5$$X$SO$_3$. We note that although most organic conductors contain metal ions in the anion complex, this system contains only organic anions, where $X$ = $CH_2CF_2$, CHF, $CF_2$, and other C,H, and F combinations at the $X$ site[1]. The influence of $X$, combined with the of β– type stacking of the cation BEDT-TTF (hereafter "ET"), leads to a variety of possible ground states. β''-$(ET)_2SF_5CH_2CF_2SO_3$ is a 5 K superconductor which has been extensively studied in high magnetic fields[3-5], and β''-$(ET)_2SF_5CHFSO_3$ shows a weak metal-to-insulator transition[6] at 6 K. Their temperature and magnetic field dependencies (for two field directions) are shown in Fig. 1.

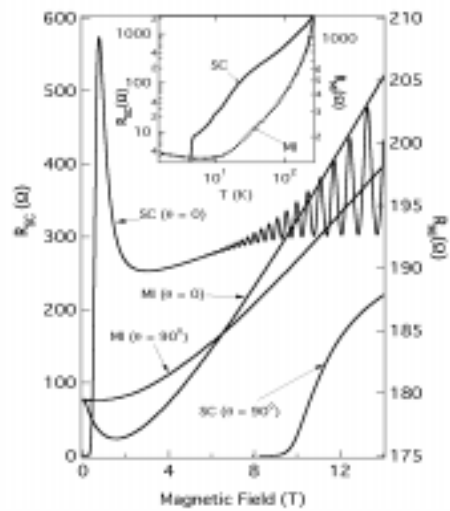

Fig. 1. Temperature dependence (inset) and magnetic field dependence at 1.6 K of the inter-plane resistance of the superconductor. β''-$(ET)_2SF_5CH_2CF_2SO_3$ (labeled SC) and the low temperature insulator β''-$(ET)_2SF_5CHFSO_3$ (labeled MI). The magnetic field dependence for both perpendicular (θ = 0) and parallel (θ = 90°) directions to the ET planes is shown. The quantum oscillations for SC(θ = 0) have a frequency of 199 T.

The second material we consider is τ-(P-$S$, $S$-DMEDT-TTF)$_2$(AuBr$_2$)(AuBr$_2$)$_y$ (where y ≈ 0.75) which is characterized by a large negative magnetoresistance (MR), and other peculiar anomalies which may indicate some sort of magnetic ordering[7,8]. Some of these features are shown in Fig. 2. Band structure for this material predicts a single, star-shaped Fermi surface[7]. Recently quantum oscillations in the resistance (Shubnikov-de Haas oscillations – "SdH") have been observed[9] above 20 T, but the SdH results show two FS closed orbit sections, both of which are less than those predicted from band calculations. The extremal Fermi surface area is expected to vary with the concentration of the (AuBr$_2$)$_y$ complex, which resides in between the τ-(P-$S$, $S$-DMEDT-TTF)$_2$(AuBr$_2$) layers, but this cannot explain the observed differences. We note also that the intervening (AuBr$_2$)$_y$ layer, and the asymmetric donor, promote a very large unit cell dimension (68 Å) along the low conductivity direction, perpendicular to the conducting layers. This may give rise to some of the unusual features.

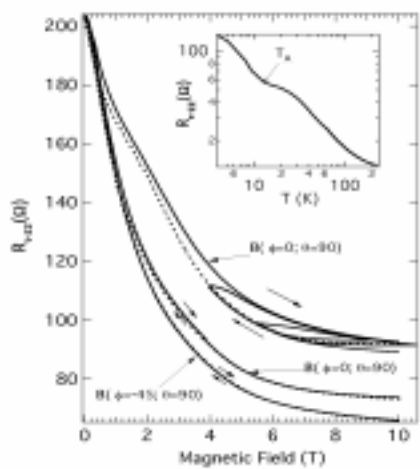

Fig. 2. Temperature dependence (inset) and magnetic field dependence at 1.6 K of the inter-plane resistance of $\tau$-(P-*S, S*-DMEDT-TTF)$_2$ (AuBr$_2$)(AuBr$_2$)$_{0.75}$. $T_A$ ( inset) refers to a well-established resistance anomaly near 12 K of unknown origin. The magnetic field dependence shows the largest hysteresis for the inter-planar ($\theta$=0) direction. Here hysteresis cycles are shown which suggest that the low resistance state is achieved by first going to higher magnetic fields.

EXPERIMENT

The method employed in the present work is that of angular dependent magnetoresistance , AMR (or AMRO in the case of angular dependent oscillations) which is unlike the Shubnikov-de Haas measurements where the magnetic field is swept at constant temperature. AMR is performed at fixed field and temperature, where the sample orientation (usually) is varied in both polar ($\theta$) and azimuthal ($\phi$) angular directions. Here $\theta$ is defined as the angle between the magnetic field and two-dimensional organic cation layer, and $\phi$ is the angle between one in-plane crystallographic axis and the in-plane projection of the magnetic field direction. In the cases studied here we find three distinctly different limiting cases of the sort of information that an AMR experiment can yield on the electronic structure of an anisotropic organic conductor.

    Typically, a current (between 1 and 10 µA below 100 Hz) is applied in a four-terminal (10 µm gold wire and gold paint) ac configuration along the least conducting crystallographic direction ($R_{zz}$), and the samples are aligned as shown in Figs. 3-5. A two-angle rotator in a 14 T superconducting magnet at the Tsukuba Magnet

Laboratory was used for the measurements. Typically θ is scanned from –110 to 110 degrees in increments of 1 degree or less. After each θ scan, ϕ is incremented in a step of 10 degrees or less and a new θ scan is made. The range of ϕ is also typically -110 to 110 degrees. An accuracy of about 0.1 degrees can be maintained for both coordinates.

RESULTS

We first report the AMRO results from the superconductor β''-(ET)$_2$SF$_5$ CH$_2$CF$_2$SO$_3$ as shown in Fig. 3. The data are presented in several different ways. First is the traditional R vs. θ plot for two different values of fixed ϕ. We see the appearance of the superconducting transition near ± 90°, and two series of oscillations, one set (the SdH effect) near θ =0°, and the others (the AMRO) which develop as θ approaches ± 90°. The superconducting critical field[4] and the SdH oscillations[3,5] depend on the in-plane magnetic field Bcos(θ) due to the layered, highly two-dimensional nature of the material. The AMRO effect, on the other hand, is dominated by the relationship[10] $\tan(\theta) = [\pi(N\pm1/4) + \mathbf{k}^{max}\cdot\mathbf{u}]/ k_b^{max}c$. Here, for a quasi-two-dimensional Fermi surface in a layered conductor with inter-planar spacing c, the inter-plane resistance (for fixed magnetic field) will become a maximum periodically in tan(θ) when N is an integer. When the Fermi surface is a symmetric cylinder, the relation above is simply $\tan(\theta) = \pi(N\pm1/4)/ k_Fc$, and such a measurement, where the period is determined by plotting the data vs. tan(θ), yields the Fermi momentum with only c as input. For more complicated cases where the in-plane hopping **u** is anisotropic, and the Fermi surface is elliptical, it is the value $k_b^{max}(\phi)$ which is determined from the experiment, and a geometric construction involving the azimuthal angle ϕ is needed[10] to determine $k_F(\phi)$. To better visualize the nature of the electronic structure near the point of minimum Fermi momentum, we carried out a local, high precision AMRO scan where Δϕ = 3°. These results, together with the Δϕ = 10° results, are shown as a contour plot for θ vs. ϕ. This presents a tomographic image of the electronic structure. Further work is underway to extract detailed information about the Fermi surface topologies of low-dimensional organic metals from tomographic information of this kind.

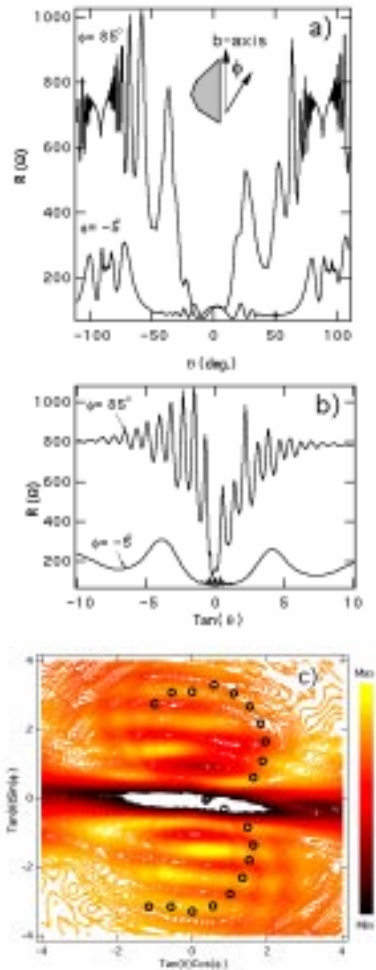

Fig. 3. a) AMRO signal from β''-(ET)$_2$SF$_5$CH$_2$CF$_2$SO$_3$ at 1.6 K and 14 T for two extreme crystallographic rotation axes vs. polar angle θ. Sample configuration is indicated. Note the signature of the critical field at 90°, and the SdH oscillations near 0°. b) Same data showing periodicity in the tan(θ). Upper curve is offset by +50 Ω to distinguish the SdH signals. c) Tomographic image of electronic structure based on the full θ–φ data set at 10 T. The periodic ridges are the loci of the amro peaks in a numonic representation. The open circles show, in azimuthal coordinates, the period of the AMRO oscillations. From this representation, a very narrow, elliptical Fermi surface is deduced by a geometrical construction [3]. Although the resulting Fermi surface topology is not as is expected from band structure, the over all method shown here is highly representative of the nature of AMRO in systems with quasi-two-dimensional warped Fermi surfaces.

The AMR and tomographic data shown in for β''-(ET)$_2$SF$_5$CHFSO$_3$ in Fig. 4 (see also Fig. 1) suggests that only the field normal to the β''-(ET)$_2$ planes removes the MI state, which otherwise remains for purely in-plane magnetic field. Although there is no apparent in-plane anisotropy in the AMR at low fields, and a minimum in the resistance for normal field (near θ = 0), for higher fields a pronounced maximum appears. Anisotropy develops both in-plane for θ = 0, and also where a maximum in the tomograph appears, for the magnetic field near φ =0, where the θ direction of the field is along the inclined side facet of the crystal. These novel effects are under further investigation.

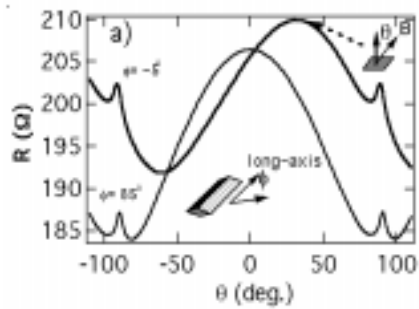
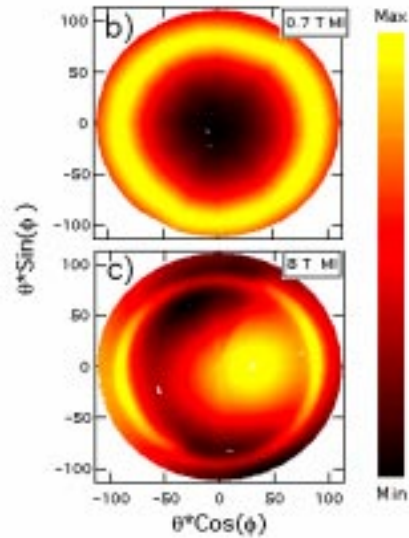

Fig. 4. a) The AMR signal from the metal-insulator system $\beta''$-$(ET)_2SF_5CHFSO_3$ at 1.6 K and 8 T. b) and c). Tomographic images at low fields in the negative MR regime and at high fields in the positive MR regime. Fig 4 a and c are from the same data set, and show the high degree of asymmetry which develops at high fields, particularly along the crystallographic direction shown in a) for $\phi$ near zero and $\theta \approx 40°$.

Turning next to the tau-phase system $\tau$-(P-$S$, $S$-DMEDT-TTF)$_2$ (AuBr$_2$) (AuBr$_2$)$_y$, the AMR reveals some striking phenomena, as shown in Fig. 5. Also, since SdH studies have shown on the same sample that there is a quasi-two-dimensional Fermi surface, behavior very similar to that observed in Fig. 3 was expected. Indeed, the periodicity of the AMRO, based on the SdH frequency (516 T) and the inter-plane unit cell dimension (68 Å), should be of order $\pi/k_Fc = 0.37$. But no AMRO have yet been observed. Second, unlike Fig. 3, the magnetoresistance is a maxium near $\theta = 0$, and not a minimum, which is at odds with the simple warped two-dimensional Fermi surface picture. Third, based on our systematic AMR study, the electronic structure (i.e. its symmetry) appears to be magnetic field dependent, as the sequence of tomographic images show in Fig. 5. At this stage, we speculate that some of the unusual properties of the tau-phase system may involve inter-plane incoherent transport mechanisms, as recently described theoretically
.

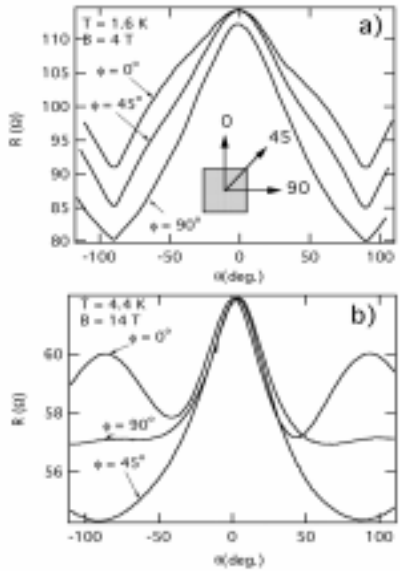
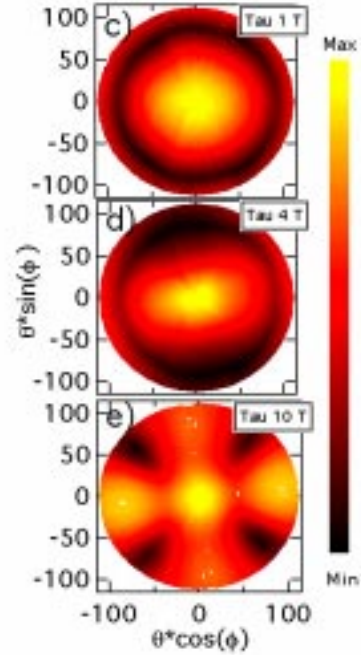

the in-plane symmetry changes from two- to four-fold, which correlates with the large decrease in resistance over the same range of field.

Fig. 5. Field dependent changes in the electronic structure of τ-(P-*S, S*-DMEDT-TTF)$_2$(AuBr$_2$) (AuBr$_2$)$_{0.75}$ with increasing magnetic field at 1.6 K. (See Fig. 2 for corresponding points in field). In a) and b) the data represent the variation of the interplane resistance R$_{zz}$ with polar angle θ for rotation of the sample along specific axes defined by the azimuthal angle. In c),d), and e) respectively, we see how the tomographic representation changes for increasing magnetic field. Here

## SUMMARY DISCUSSION

We have shown, in this very brief report of on-going investigations, that highly systematic studies of the inter-plane magnetoresistance of novel low dimensional materials reveals tomographic information about their electronic structure. This not only involves the standard methods to determine the azimuthal shape of a quasi-two-dimensional Fermi surface, but it also brings new insight into systems where other anisotropic, or anomalous magnetic field dependent electrical transport mechanisms are at work.

ACKNOWLEDGEMENTS

This work is supported by NSF-DMR-99-71714. Measurements were carried out at the Tsukuba Magnet Lab. Work at Argonne National Laboratory is sponsored by the US DoE, OBES, Div. Mat. Sci., under contract W-31-109-ENG-38. Work at Portland State Univ. is supported by NSF - CHE-9904316 and the Petroleum Research Fund ACS-PRF34624-AC7.